\documentclass[preprint,showpacs,showkeys,preprintnumbers,amsmath,amssymb]{revtex4}

\usepackage{natbib}
\usepackage{graphicx}
\usepackage{dcolumn}
\usepackage{bm}
\usepackage{longtable}

\begin{document}


\title{Structural, Magnetic, and Electronic Properties of the Monometallofullerene Gd@C$_{82}$: Theory}

\author{Ali Sebetci}
 \altaffiliation{Current affiliation: Department of Chemistry, University of California, Riverside, CA 92521 USA}
 \email{sebetci@ucr.edu}
\affiliation{IFW Dresden, P.O. Box 270116, D-01171 Dresden, Germany}

\author{Manuel Richter}
 \email{m.richter@ifw-dresden.de}
\affiliation{IFW Dresden, P.O. Box 270116, D-01171 Dresden, Germany}

\date{\today}

\begin{abstract}

The structural, electronic, and magnetic properties of Gd@C$_{82}$ endohedral metallofullerene have been studied 
by employing on-site correlation corrected, scalar-relativistic and 
full-relativistic
density functional theory 
within the local density and generalized gradient approximations.   
The experimentally observed reduction of the magnetic moment of Gd@C$_{82}$ with respect to that of a free Gd$^{+3}$ ion can 
be explained
by a tiny hybridization between unoccupied Gd-4f states
and carbon-$\pi$ states, resulting in a generic
antiferromagnetic coupling of the Gd-4f spin with the remaining unpaired 
spin in the hybridized molecular orbital. 

\end{abstract}


\maketitle

\section{Introduction}

Rare-earth (RE) based metallofullerenes such as RE@C$_{60}$~\cite{Bolskar}, RE@C$_{82}$~\cite{Kato}, 
and RE$_{3}$N@C$_{80}$~\cite{Fatouros} can be solved in water and functionalized with poly- and multihydroxyl 
groups in order to be used as contrast enhancing agent for magnetic 
resonance imaging (MRI). Their advantage is not only that they are safer
than the commercial MRI contrast agents such as Gd-DTPA since the toxic RE ions are totally 
encapsulated by the fullerenes and cannot escape from the cage under biological 
conditions but also that they can produce proton relaxivities nearly twenty times 
larger than those of the commercial agents~\cite{Er-Yun,Kato}. Proton relaxivity is the ability of magnetic 
compounds to increase the relaxation rates of the surrounding water proton spins. 

Metallofullerenes are also promising in photoelectrochemical
cell~\cite{Yang} and molecular memory~\cite{Slanina,Jones} applications as well as spintronics devices~\cite{Shimada}. 
Therefore, endohedral monometallofullerenes of type RE@C$_{82}$, beside the others, have attracted 
a wide interest during the last decade~\cite{Funasaka,Kessler,
Kobayashi,Huang,Suenaga,Furukawa,Nadai,Bondino,Kitaura}. Although there is a considerable
number of investigations of these systems published,
the structural and magnetic properties of monometallofullerenes  
and the details of the interaction between the metal atom and the carbon cage still need to be 
clarified. 

The cage structures of Sc@C$_{82}$~\cite{Nishibori1} and La@C$_{82}$~\cite{Nishibori2} have been determined 
to have $C_{2v}$ symmetry
by a synchrotron radiation powder diffraction based structural analysis using the maximum entropy method (MEM).
Inside the carbon cages, the RE atoms are located at an off-centre position adjacent to a carbon six-membered ring. 
This structure has been confirmed by $^{13}$C NMR spectroscopy~\cite{Akasaka}.
The similarity in UV/vis/NIR spectra of 10 kinds of RE@C$_{82}$ 
(RE = La, Ce, Pr, Nd, Gd, Tb, Dy, Ho, Er, and Lu)~\cite{Akiyama} 
dissolved in toluene strongly suggests that Gd@C$_{82}$ possesses $C_{2v}$ symmetry as 
Sc@C$_{82}$ and La@C$_{82}$. An extended X-ray absorption fine structure (EXAFS) study has proposed a position
of the Gd ion in the C$_{82}$ cage above the carbon hexagon~\cite{Geifers}. However, in a later 
experimental study, Nishibori {\it et al.}~\cite{Nishibori3} have 
suggested, on the basis of synchrotron radiation powder structure MEM analysis, 
that the Gd atom 
would be located in the vicinity of the C-C double bond on the same $C_{2}$ molecular axis of the C$_{82}$ cage, 
but opposite to the six-membered ring where Sc and La atoms in Sc@C$_{82}$ and La@C$_{82}$ 
are known to be placed. On the theoretical side, it has been reported in an earlier  
density functional theory (DFT) calculation~\cite{Kobayashi} 
that C$_{82}$-$C_{2v}$ cage symmetry is the most stable structure for La@C$_{82}$ 
metallofullerenes where La ions are strongly bonded to the hexagonal rings of the cages. 
In this study by Kobayashi and Nagase~\cite{Kobayashi}, 
structural relaxation of Gd@C$_{82}$ metallofullerene with different initial
positions 
of the Gd atom in the cage was not considered. Later, Senapati {\it et al.}~\cite{Senapati,Wang,Senapati2} 
have reported that their scalar relativistic DFT calculations with effective core potentials (ECP) did 
not agree with the MEM/Rietveld-based X-ray synchrotron powder diffraction structure of Nishibori 
{\it et al.}~\cite{Nishibori3}. They have found the most stable position of the Gd atom adjacent to the C-C 
double bond but not on the $C_{2}$ molecular axis of the C$_{82}$ cage. Finally, the disagreement 
on the position of the Gd ion in the cage has been solved both theoretically and experimentally:
Mizorogi and Nagase~\cite{Mizorogi} performed DFT calculations and revealed that the so-called anomalous
structures with Gd close to the double bond are unstable and do not correspond to energy minima, and Liu {\it et al.}~\cite{Liu}
have shown by an X-ray absorption near-edge structure (XANES) study that the Gd ion lies above the hexagon
on the $C_{2}$ axis, like La and Sc.

Effective magnetic
moments $\mu_{\rm eff}$ 
of RE@C$_{82}$ metallofullerenes have been measured by
employing Superconducting Quantum Interference Device (SQUID) magnetometers for 
RE = La, Gd~\cite{Funasaka}, 
RE = Gd, Tb, Dy, Ho, Er~\cite{Huang}
and by soft X-ray magnetic circular dichroism (SXMCD) spectrometers 
for RE = Gd, Dy, Ho, Er~\cite{Kitaura}.
It has been found that they are significantly smaller than those of 
the corresponding free RE$^{3+}$ ions. The amount of the reduction in effective magnetic moment is different
for each metallofullerene with a general trend that the higher the orbital momentum, the larger the magnitude
of the reduction~\cite{Huang}. Particularly, the measured values of 6.90 $\mu_{B}$~\cite{Funasaka}, 
6.91 $\mu_{B}$~\cite{Huang}, and 6.8$\pm$0.5 $\mu_{B}$~\cite{Kitaura} for Gd@C$_{82}$, which were obtained 
by fitting the experimental data to the Curie-Weiss law, correspond to an approximately 13\% reduction
in the effective moment compared to the theoretical value of 7.94 $\mu_{B}$ of the free trivalent Gd ion.
The case of Gd is simpler than that of other RE, since the Gd magnetic moment is almost completely spin-dominated. 
Thus, the effective moment can be related to the allowed spin multiplicities, M = 2S+1 
which is even for Gd$^{+3}$ and
odd for Gd@C$_{82}$. Indeed, $\mu_{\rm eff} = 7.94$ $\mu_{B}$ corresponds to M = 8 (Gd$^{+3}$) while
$\mu_{\rm eff}=6.93$ $\mu_{B}$ would correspond to M = 7 if a vanishing orbital contribution is presumed.
Senapati {\it et al.}~\cite{Senapati} have calculated the 
total energy difference between different spin multiplicities of Gd@C$_{82}$ resulting from 
ferromagnetic (M = 9) 
and antiferromagnetic (M = 7)
arrangements of the Gd f-electrons and the remaining odd electron, and concluded that 
M = 7 is the ground state. This agrees with the experimental result by Furukawa 
{\it et al.}~\cite{Furukawa} who estimated the energy 
difference between the two states 
to be 1.8 meV. 

Later, however,
using the GAUSSIAN 03 code and a hybrid exchange correlation functional (B3LYP),
Mizorogi and Nagase~\cite{Mizorogi}
have found that the M = 9 state is 0.4 meV more stable than M = 7 state.
This is in contradiction with experiment.

The latter authors argued
that Senapati {\it et al.}~\cite{Senapati}
obtained their conclusion only for one of the anomalous positions~\cite{Wang,Senapati2}. 
In addition, we note that in all previous calculations on Gd@C$_{82}$
the Gd-4f-electrons were treated without any on-side correlation correction. 
Such corrections are obligatory for a decent description
of 4f states in most rare-earth elements~\cite{Richter98}
and have been applied, e.g., to the 4f states of Gd@C$_{60}$~\cite{Sabirianov07} recently.

In this work, our aim is to investigate the origin of the
M = 7 ground state of Gd@C$_{82}$ 
theoretically.
Starting from local spin density (LSDA) and generalized gradient approximations (GGA),
we include
on-site correlation correction, spin-orbit coupling, and non-collinearity effects
in the DFT
calculations which were not considered in 
any of the above mentioned previous theoretical approaches. 

\section{Computational Details}

Three DFT codes have been used in our investigation: NWChem~\cite{NWChem}, FPLO-7.28~\cite{FPLO}, and OpenMX~\cite{OpenMX}.
This has become necessary as none of the codes includes all technical
prerequisites to solve the problem. 

Scalar-relativistic geometry optimizations without any on-site correlation
correction have been performed with the program package NWChem
to compare with results given in the literature.
Then, the effect of on-site correlation corrections on the geometry
has been studied by OpenMX. It turns out, that both methods give similar
results but the OpenMX geometry data slightly deviate from the experimentally observed $C_{2v}$
symmetry. Thus, we used the NWChem geometry data in the further calculations
despite the fact that they were obtained by calculations without
on-site correlation corrections.

The magnetic ground state has been investigated using
on-site correlation corrected calculations with the FPLO code
in scalar-relativistic mode.
We also carried out the final analysis of the electronic
structure with this code in order to clarify the origin of the
observed antiferromagnetic coupling.
In addition, the effect of spin-orbit coupling and non-collinear
magnetic moments has been checked with the OpenMX code.

In the calculations with NWChem, the hybrid 
functional B3LYP~\cite{Becke} has been chosen as the exchange-correlation
functional, since this functional is known to provide geometries
close to experiment in carbon systems. 
The scalar-relativistic
effective core potential (ECP) and basis set developed by Cundari and Stevens~\cite{Cundari} were used for Gd
where the 46 inner electrons are replaced by the ECP and the outer 
4f$^{7}$5s$^{2}$5p$^{6}$5d$^{1}$6s$^{2}$ electrons are treated in the valence region.
The split-valence d-polarized 3-21G* basis set was used for C.  

The program package OpenMX~\cite{OpenMX} is based on norm-conserving
pseudopotentials~\cite{Troullier} and pseudo-atomic localized basis functions. 
In the calculations with OpenMX, the same
outer electrons of the Gd atom (as in the NWChem calculations) were 
treated as valence electrons in the self consistent field iterations. The pseudo-atomic orbitals
have been constructed by minimal basis sets (two-s, one-p, one-d, and one-f for Gd, and 
one-s, and one-p for C) within 8.0 Bohr
radii 
cutoff radius of the confinement potential 
for Gd and 5.0 Bohr 
radii 
for C. 
The GGA+U approach~\cite{Han} was applied
in the atomic limit version in order to describe the correlated
behavior of the Gd-4f shell.
A value of U-J = 7.2 eV was used and the GGA functional
was parameterized according to PBE~\cite{Perdew}. 
A similar value, U = 7.6 eV, was recently used in a similar calculation for Gd@C$_{60}$~\cite{Sabirianov07}.
The cutoff
energy for the real-space grid integration in the construction of the density matrix elements~\cite{Ozaki2} 
has been chosen as 500 Ryd.
The convergence criteria chosen were 0.1 $\mu$Ha for the total
energy and 0.1 mHa/Bohr for the geometry optimization. 

The program FPLO is a full-potential
all-electron local orbital code.
It employs a fixed orbital basis with 4f, 5s5p5d5f, 6s6p6d, 7s valence
orbitals for Gd and 1s, 2s2p, 3s3p3d valence orbitals for C.
The LDA+U approach in the atomic limit version was applied
with different values of U  and J = 0.8 eV.
The LDA functional was parameterized
according to Perdew and Zunger~\cite{Zunger}. 

\section{Results and Discussion}

\subsection{Geometric Structure} 

Let us recall: 
It is reported in a powder diffraction based structural 
analysis~\cite{Nishibori3}
that the endohedral structure of
Gd@C$_{82}$
is anomalous, namely it is not the same as previously determined structures of La@C$_{82}$ and Sc@C$_{82}$. 
In addition, not the same anomalous structure suggested by the experiment but yet another 
anomalous structure has been predicted by DFT calculations~\cite{Senapati,Wang,Senapati2}.
However, most recently Liu {\it et al.}~\cite{Liu} have shown by XANES that Gd@C$_{82}$ has a normal
endohedral structure. Furthermore, Mizorogi and Nagase~\cite{Mizorogi} revealed by DFT calculations that the anomalous 
structures are unstable. 

\begin{figure}
\includegraphics[scale=0.8]{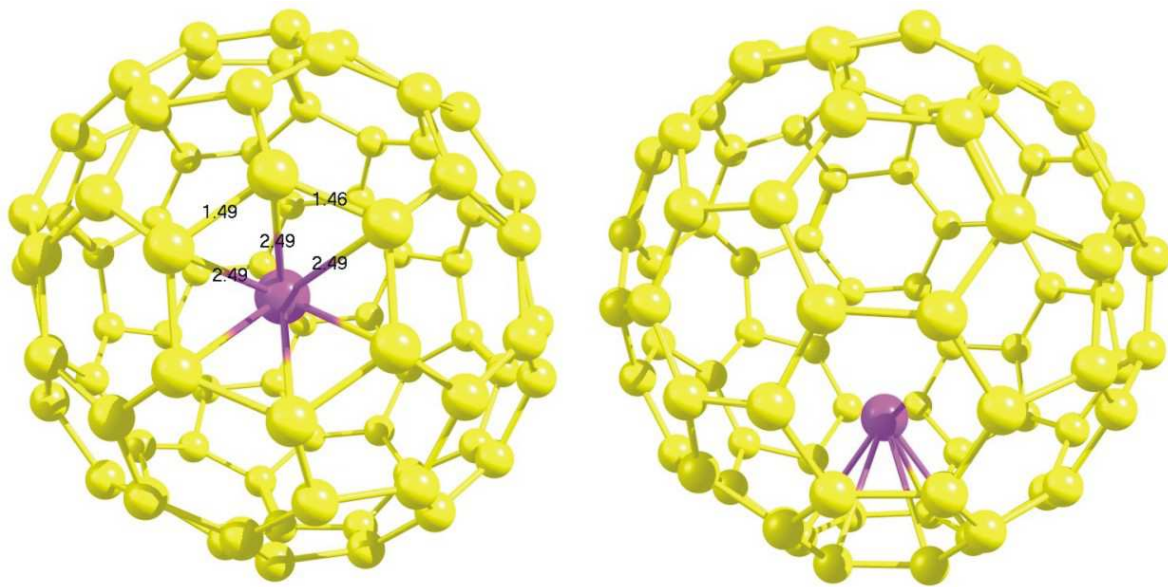}
\caption{\label{Structure}(Color online) Top and side views of
Gd@C$_{82}$ (relaxed structure). Distances are given in ${\rm\AA}$.}
 
\end{figure}

Our calculations support the findings of the most recent theoretical and experimental 
investigations. We have optimized the geometry of 
Gd@C$_{82}$ by NWChem and obtained 
a structure with $C_{2v}$ symmetry
(see Fig.~\ref{Structure}) where the Gd atom sits at the centre of one of the hexagonal carbon rings. 
The optimized
coordinates of the inequivalent atoms can be found in Table~\ref{coord}.
The calculated Gd-C bond length is 2.49 ${\rm\AA}$, while C-C bond lengths amount to 1.46 and 1.49 ${\rm\AA}$.
In Ref.~\cite{Mizorogi},
using the same exchange-correlation functional,
the Gd-C bond length was calculated as 2.47 ${\rm\AA}$. OpenMX optimization by GGA+U 
yields a very similar
structure where the Gd atom locates at a position slightly off the centre of the ring. OpenMX calculations also
reveal that the previously suggested two anomalous geometries have nearly 1.74 eV higher energies than the ground 
state structure. 

\subsection{Magnetic Properties}

The magnetic moment of the unpaired electron in the hybrid orbital can couple with that of the seven Gd-4f electrons 
either ferrromagnetically or antiferromagnetically. 
While the parallel arrangement corresponds to M = 9 for the metallofullerene, 
the antiparallel arrangement results in M = 7. The measured values of the effective magnetic moment 
of Gd@C$_{82}$ correspond to the M = 7 state. Furukawa {\it et al.}~\cite{Furukawa} have determined 
experimentally that the antiferromagnetic arrangement has 1.8 meV lower energy than the 
ferromagnetic one. On the other hand, Curie-Weiss law fitted experimental data by Funasaka {\it et al.}~\cite{Funasaka} and
by Huang {\it et al.}~\cite{Huang} suggest that the antiferromagnetic arrangement is stable up to room temperature.
   
The inclusion of on-site correlation effects is mandatory for
a correct description of the Gd-4f states
and their influence on magnetic properties~\cite{Richter98}.
We have calculated the energy differences between the two magnetic states
in LSDA+U approximation, using the FPLO code.
Reasonable values of U for 4f states of neutral rare-earth atoms
range between 6 eV and 7 eV~\cite{Richter01}.
These values are expected to be somewhat enhanced in a cationic
situation due to related orbital contraction.
Therefore, we considered the range U = 8, 10, and 12 eV. 
All considered values yield an M = 7 ground state, in accordance with the
experiment.
The M = 9 state lies 9 (3, 0.3) meV higher in energy for U = 8 (10, 12) eV.
The related spin moments for all inequivalent atoms 
are given in Table~\ref{coord}. 

The influence of spin-orbit coupling and non-collinear
spin arrangement on the relative positions of M = 7 and M = 9
states has been checked with the OpenMX code.
Both effects produce only marginal changes and can be neglected.

\subsection{Electronic Structure}

\begin{figure}
\includegraphics[scale=0.8]{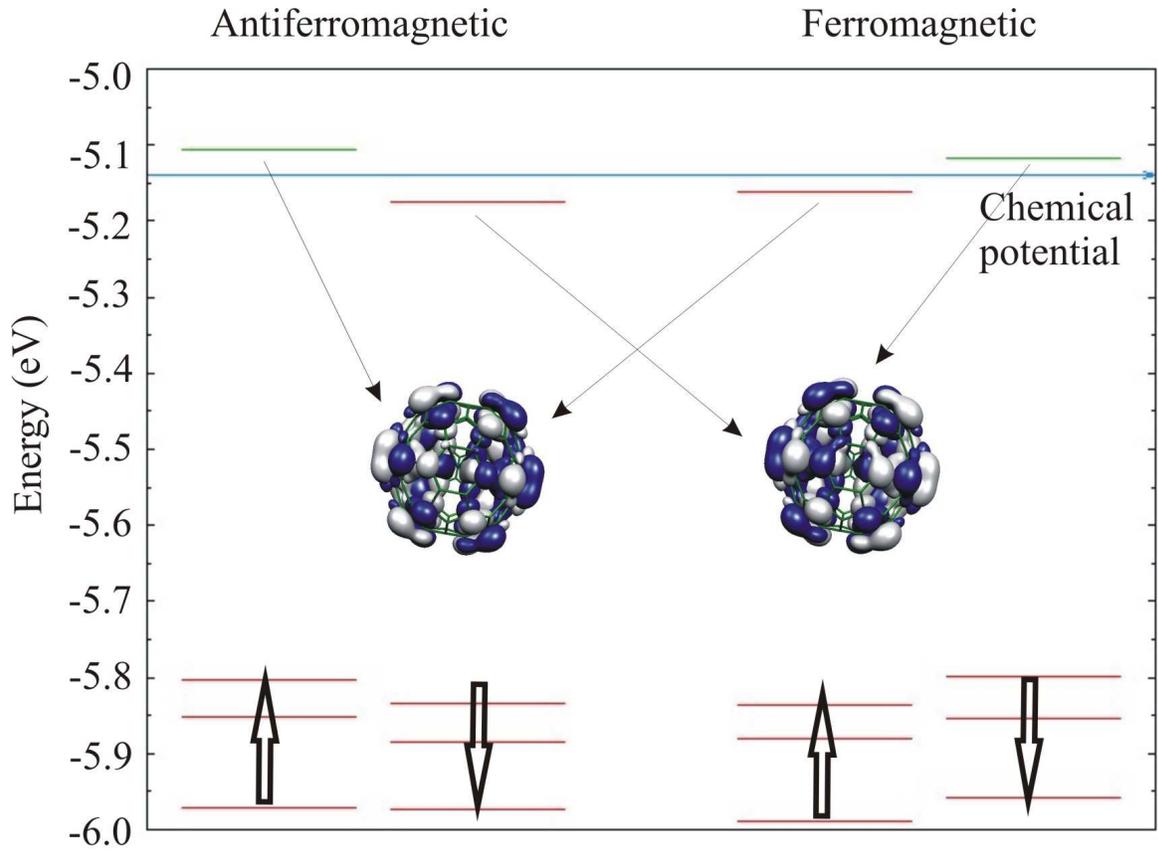}
\caption{\label{EL}Single electron energy levels of Gd@C$_{82}$ for the antiferromagnetic
(M = 7) and ferromagnetic (M = 9) 
arrangements, obtained by scalar-relativistic LSDA+U calculations using the FPLO code, U = 8 eV.
Dark (red online) and light lines (green online) represent occupied and unoccupied electron energy levels, respectively.
The dark (blue online) and the light (gray online) areas in the density pictures represent positive and negative values
of the 
related HOMO and LUMO
wave function. 
It is hard to see the Gd contribution on these pictures since it amounts to about 1.5\% only.} 
\end{figure}

The electron energy levels of the Gd@C$_{82}$ metallofullerene
close to the chemical potential, obtained by the FPLO code in
scalar-relativistic mode and using LSDA+U, U = 8 eV,
 are shown in Fig.~\ref{EL} for M = 7 and M = 9.
There is one almost spin-degenerate level at the chemical
potential, well separated from the next lower occupied
and next higher unoccupied levels by about 
0.6-0.7 eV. 
In the antiferromagnetic M = 7 ground state, the HOMO is in the
spin-down channel and the LUMO in the spin-up channel.
The opposite situation is found in the ferromagnetic M = 9 state.

The occupied 4f levels (spin-up by definition) are
situated at $\sim$ -16 eV, about 11 eV below the chemical 
potential and outside the displayed energy range in
Fig.~\ref{EL}. Thus, they are chemically inert and only
contribute a spin magnetic moment.
The calculated position of the occupied 4f levels agrees with the recent
photoemission data of Gd@C$_{60}$ locating the 4f emission
at a -10 to -11 eV binding energy~\cite{Sabirianov07}.
The unoccupied 4f-spin-down levels are separated from
the 4f-spin-up levels by the sum of exchange splitting (about 5 eV
in the Gd-4f shell) and the term U-J.
Thus, they are situated at $\sim$ -4 eV, about 1 eV above the 
chemical potential.

The 6s and 5d electrons of the Gd atom essentially
occupy empty states of the carbon cage, since the 
chemical potential of the empty cage lies below the 
Gd chemical potential. However, Mulliken population analysis
shows a Gd occupation of $4f^{7.02}5d^{1.16}6s^{0.07}$ in the
metallofullerene. The 5d occupation indicates a non-negligible
degree of covalency. Indeed, population analysis of the
individual $\pi$-like molecular orbitals reveals a 5d contribution
in the percent range. 

What might be more interesting is
a 4f contribution of about 1 percent to the (spin-down)
HOMO of the M = 7 ground state
and to the (spin-down) LUMO of the M = 9 state.
Such a contribution is not present in the spin-up
channel (LUMO of the M = 7 ground state and HOMO
of the M = 9 state), since the 4f-spin-up states are
inert. This difference provides an explanation
for the observed and also calculated lower energy of the M = 7
state: Due to the position of the unoccupied 4f-spin-down
states close to the chemical potential, more variational
freedom is available for the spin-down molecular orbitals
and their energy is reduced with respect to the
spin-up levels, thus favoring a spin-down HOMO and
a related antiferromagnetic coupling between the
Gd-4f shell and the unpaired electron at the cage.
Using a simplifying two-level model,
one can roughly estimate the lowering of the spin-down 
level closest to the chemical potential by interaction
with the 4f states:

\[
\varepsilon_{\rm hyb}^{\downarrow} - \varepsilon_{\pi} \approx
- \frac{(\varepsilon_{\rm 4f}^{\downarrow} - \varepsilon_{\pi})^2 |C_{\rm 4f}^{\downarrow}|^2}{\varepsilon_{\rm 4f}^{\downarrow} - \varepsilon_{\pi}}
\]

Here, $\varepsilon_{\rm hyb}^{\downarrow}$ and $\varepsilon_{\pi}$
denote the level of the 4f-$\pi$ hybrid and the pure $\pi$ state,
respectively, $\varepsilon_{\rm 4f}^{\downarrow}$ denotes the
position of the empty 4f state, and $|C_{\rm 4f}^{\downarrow}|^2$
denotes the squared 4f contribution to the eigenvector of
the hybrid state.
The latter amounts to about 1\% for U = 8 eV. Using
$\varepsilon_{\rm 4f}^{\downarrow} - \varepsilon_{\pi} \approx 1$ eV,
we arrive at a level lowering in the order of 10 meV,
accounting for the calculated energy difference between M = 7 and M = 9 states, 9 meV.
This energy difference is reduced, if U is enhanced but stays
positive up to the (unreasonably high) value of U = 12 eV.

A final remark is in place to explain the relative
position of HOMO and LUMO in the ferromagnetic state.
If M = 9 is forced in the calculation, the spin-up
molecular orbital at the chemical potential is occupied.
The related spin density (mainly situated on the carbon
cage, see inset of Fig.~\ref{EL}) creates an exchange field
that lowers the position of the spin-up level.
The same happens in the M = 7 case for the spin-down level.
Both shifts are due to (unphysical) self-exchange of the
unpaired electron on the cage.
In the latter case, the spin-down level was anyway lower
in energy than the spin-up level due to 4f-$\pi$ hybridisation, and self-exchange
enhances the splitting between the two levels. 
In the M = 9 case,
self-exchange and hybridisation effects compete
and the level splitting is lower than in the M = 7 case.
Though the total energy is slightly shifted by the described
effect in an unphysical manner, this shift is virtually
equal in the M = 7 and the M = 9 state and hence does
not influence the energy difference between these states.

\section{Conclusions}

In the present study, the structural, magnetic, and electronic properties of Gd@C$_{82}$ endohedral metallofullerene have been 
investigated by
different approximations within density functional theory. 
It is confirmed that the lowest energy structure 
of Gd@C$_{82}$ has $C_{2v}$ symmetry where the Gd atom is located at 
a position on the symmetry axis, adjacent to a carbon six-membered ring. 
The highest molecular orbitals of Gd@C$_{82}$ are not pure $\pi$ states but 
(4f)-d-$\pi$ hybridized molecular orbitals.
The experimentally observed reduction of the Gd@C$_{82}$ magnetic moment with respect to that of a free Gd$^{+3}$ ion
is due to
antiferromagnetic coupling between the 4f electrons of the Gd atom and the remaining unpaired 
electron on the hybridized molecular orbital.
The reason for this antiferromagnetic coupling is a small
hybridization of the unoccupied 4f-spin-down states
with the carbon $\pi$-states. It yields an M = 7 ground state that should be
generic for all Gd-carbon systems with unpaired electrons.
For example, an M = 7 ground state has also been found for Gd@C$_{60}$ in a recent
calculation~\cite{Sabirianov07}.

\begin{acknowledgements}
We like to thank Alexey Popov, Klaus Koepernik, Helmut Eschrig,
Gotthard Seifert, and
Lothar Dunsch for discussions.
Financial support has been provided by the German Science Foundation
via SPP 1145.
\end{acknowledgements}


\begingroup
\squeezetable
\begin{table}
\caption{\label{coord}Coordinates and 
spin moments for all inequivalent atoms of Gd@C$_{82}$.
The coordinates originate from geometry optimization
using NWChem with the B3LYP functional. The moments were obtained
at the given coordinates by
FPLO with LSDA+U using the PZ-81 functional and U = 8 eV, J = 0.8 eV.}
\begin{ruledtabular}
\begin{tabular}{l c c c c c c }
 Number& Atom    & \multicolumn{3}{c}{Coordinates (${\rm\AA}$)}   & \multicolumn{2}{c}{Spin Moments ($\mu_{B}$)} \\
       \cline{3-5}                    \cline{6-7}
       &         & X       & Y          & Z         & M = 9   & M = 7 \\
 \hline
    1  & Gd      & 0.000   & 0.000 & 1.822          &  7.143  &  7.119        \\
    2  & C       & -1.215  & 3.943 & 0.643          &  0.002  & -0.004        \\
    3  & C       & 0.000   & 3.853 & 1.368          &  0.000  &  0.002       \\
    4  & C       & 0.000   & 3.068 & 2.573          & -0.010  & -0.003         \\
    5  & C       & 1.213   & 4.018 & -0.789         &  0.025  & -0.028        \\
    6  & C       & 0.000   & 3.960 & -1.534         & -0.005  &  0.005      \\
    7  & C       & 2.387   & 3.182 & 1.064          &  0.032  & -0.037        \\
    8  & C       & 2.354   & 2.368 & 2.217          & -0.003  &  0.006       \\
    9  & C       & 1.160   & 2.363 & 3.026          &  0.012  & -0.022         \\
   10  & C       & 3.108   & 1.166 & 2.198          &  0.000  &  0.002       \\
   11  & C       & 2.408   & 2.669 & -2.458         &  0.025  & -0.028         \\
   12  & C       & 2.400   & 3.325 & -1.252         &  0.007  & -0.008        \\
   13  & C       & 3.091   & 2.754 & -0.121         & -0.001  &  0.001      \\
   14  & C       & 3.687   & 1.465 & -0.207         &  0.003  & -0.003         \\
   15  & C       & 3.764   & 0.720 & 1.004          &  0.025  & -0.028        \\
   16  & C       & -1.205  & 2.652 & -3.285         & -0.005  &  0.006      \\
   17  & C       & 0.000   & 3.256 & -2.829         & -0.000  & -0.000         \\
   18  & C       & 2.661   & 0.000 & 2.943          &  0.016  & -0.029        \\
   19  & C       & 1.512   & 0.000 & 3.803          & -0.000  & -0.029        \\
   20  & C       & 0.743   & 1.238 & 3.847          & -0.021  & -0.017        \\
   21  & C       & 3.097   & 1.406 & -2.599         &  0.034  & -0.037        \\
   22  & C       & 3.660   & 0.735 & -1.481         &  0.014  & -0.015        \\
   23  & C       & -1.203  &-1.426 & -4.021         &  0.035  & -0.039        \\
   24  & C       & -2.403  &-0.683 &-3.638          & -0.003  &  0.004       \\
   25  & C       & 0.000   &-0.739 &-4.330          & -0.005  &  0.006       \\
\end{tabular}
\end{ruledtabular}
\end{table}
\endgroup

\newpage

\end{document}